\documentclass[11pt]{article}
\usepackage{moriond,epsfig}

\def\be{\begin{equation}}
\def\ee{\end{equation}}
\def\bea{\begin{eqnarray}}
\def\eea{\end{eqnarray}}

\begin{document}
\vspace*{4cm}
\title{METAL ENRICHMENT AND ENERGETICS OF GALACTIC WINDS IN GALAXY CLUSTERS}

\author{ W. KAPFERER$^{1}$, D. BREITSCHWERDT$^{2}$, W. DOMAINKO$^{1}$, S. SCHINDLER$^{1}$,\\ E. VAN KAMPEN$^{1}$, S. KIMESWENGER$^{1}$, M. MAIR$^{1}$ }

\address{1-Institute for Astrophysics, University of Innsbruck,\\
6020 Innsbruck, Austria\\
2-Institute for Astronomy, University of Vienna,\\
1180 Vienna, Austria\\}

\maketitle\abstracts{ We investigate efficiency and time
dependence of metal enrichment processes in the Intra-Cluster
Medium (ICM). In this presentation we concentrate on the effects
of galactic winds. The mass loss rates due to galactic winds are
calculated with a special algorithm, which takes into account
cosmic rays and magnetic fields. This algorithm is embedded in a
combined N-body/hydrodynamic code which calculates the dynamics
and evolution of a cluster. We present mass loss rates depending
on galaxy properties like type, mass, gas mass fraction and the
surrounding ICM. In addition we show metallicity maps as they
would be observed with X-ray telescopes.}

\section{Introduction}
From X-ray spectra it is evident that the ICM contains metals
(Fukazawa et al. 1998). As heavy elements are only produced in
stars the processed material must have been ejected by cluster
galaxies into the ICM. Possible transport processes are ram
pressure stripping (Gunn \& Gott 1972), galactic winds (De Young
1978), galaxy-galaxy-interactions or jets from AGNs. In this work
we concentrate on the effects of galactic winds. The mass loss
rates due to galactic winds are calculated taking into account the
galaxies properties. The ejected material acts as an input for our
combined N-body hydrodynamic code which calculates the
redistribution of the material due to the evolution of a galaxy
cluster. The properties of the galaxies are obtained by applying
seminumerical galaxy evolution models.

\section{The combined N-body Hydrodynamic Code}

Large-scale structure formation is simulated with a $\Lambda$CDM
N-body tree code with an additional semi numerical model for
galaxy formation (van Kampen et al. 1999). The gas is treated with
a PPM (piecewise parabolic method) hydrodynamic grid code (Colella
\& Woodward, 1984). To obtain high resolution in the cluster
center the hydrodynamic code uses a fixed mesh refinement
(Ruffert, 1992). The simulations are covering a volume of 20
(h$^{-1}$) Mpc$^{3}$. In addition four nested grids each with the
same resolutions are calculated to obtain the highest resolution
in the center (Fig.1).

\begin{figure} [h]
\center \psfig{figure=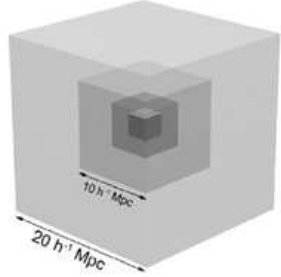, height=1.5in} \caption {The fixed mesh
refinement in our hydrodynamic simulations}
\end{figure}

\section{The Galactic Wind Algorithm}

The mass loss rate due to supernova driven galactic winds is
calculated with a code developed by Breitschwerdt et al. (1991).
The algorithm calculates for a given galaxy the mass loss rate and
wind properties like the velocity of the ejected matter as a
function of distance to the galaxy or the pressure flow. As an
input for the wind code galaxy parameters like halo mass, disc
mass, spin parameter, scale length of the components, temperature
distribution of the ISM, magnetic field strength and density
distribution as well as stellar density distribution are required.
All those properties are calculated within the semi numerical
model for galaxy evolution. In order to save computing time we
performed parameter studies. The results are sumerised in a look
up table. One example of such a table is shown in Fig. 2. The mass
loss rate is plotted as a function of the spin parameter and the
mass of the disc. In this example the halo mass of
$3\times$10$^{11}$ M$_{\odot}$ is assumed to be constant for all
model galaxies.

\begin{figure} [h]
\center \psfig{figure=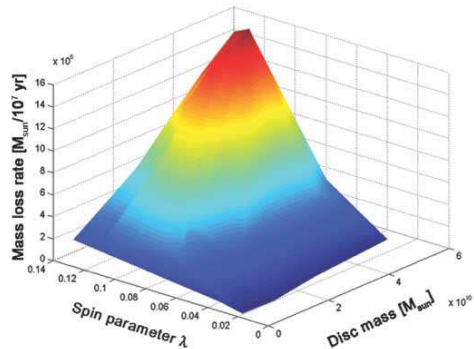, height=2in} \caption {Mass loss
rate as a function of disc mass and spin parameter for a given
constant halo mass of 3$\times$10$^{11}$ M$_{\odot}$.}
\end{figure}

The larger the disc mass and the spin parameter the higher is the
mass loss rate. The extent of the disc increases with increasing
spin parameter and therefore the probability for supernova
explosions at large radii increase as well.

\section{First Results}

We then applied the galactic wind tables to our combined N-body
hydrodynamic code with semi numerical galaxy evolution. The model
cluster consists of 1850 galaxies total of which 338 galaxies show
a wind. The overall average mass loss rate due to galactic winds
is about 25 M$_{\odot}$/yr. Note that starburst galaxies were not
taken into account. The mass loss rates together with the
metallicities of the galaxies act as an input for the hydrodynamic
code. The code calculates the dynamics of the ICM and hence can
simulate mixing of the ejected matter and the ICM due to a cluster
merger process. In figure 3 the temperature distribution projected
along the line of sight the model cluster is shown before and
after the cluster merger. Due to the merging process several
outgoing shockwaves can be seen. The time between the two images
is about 4.9 Gyrs in a $\Lambda$CDM cosmology, with
$\Omega_\Lambda=0.7$, $\Omega_{\rm m}=0.3$, $h_0=0.7$, and
$\sigma_8=0.93$.

\begin{figure} [h]
\psfig{figure=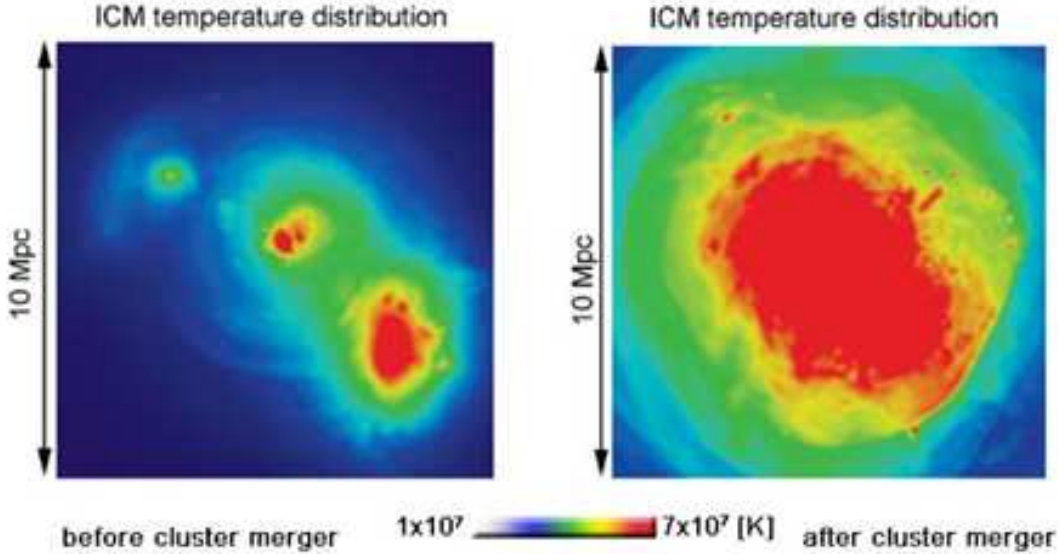, height=3.2in} \caption {Projected
temperature distribution of our simulation cluster. Left image
before merger right image after merger. $\Delta$t=4.9 Gyrs.}
\end{figure}

In figure 4 the ejected matter due to galactic winds and the
mixing with the ICM due to the merger is shown. In the upper panel
a 3D distribution of the metallicity of the ICM in a 10 Mpc$^{3}$
cube is displayed, that has been accumulated since z=1 by galactic
winds. In the central 5 Mpc$^{3}$ cube the mixing of the ejected
material with the ICM is more efficient than in the outer regions
of the cluster. As the density of the ICM increases towards the
cluster center the winds will be suppressed within a region of
several 100 kpc around the center. In the outer regions of the
cluster where the density decreases galactic winds are more
efficient to enrich the ICM than ram pressure stripping (see
Domainko et al. this volume).

\begin{figure} [h]
\center \psfig{figure=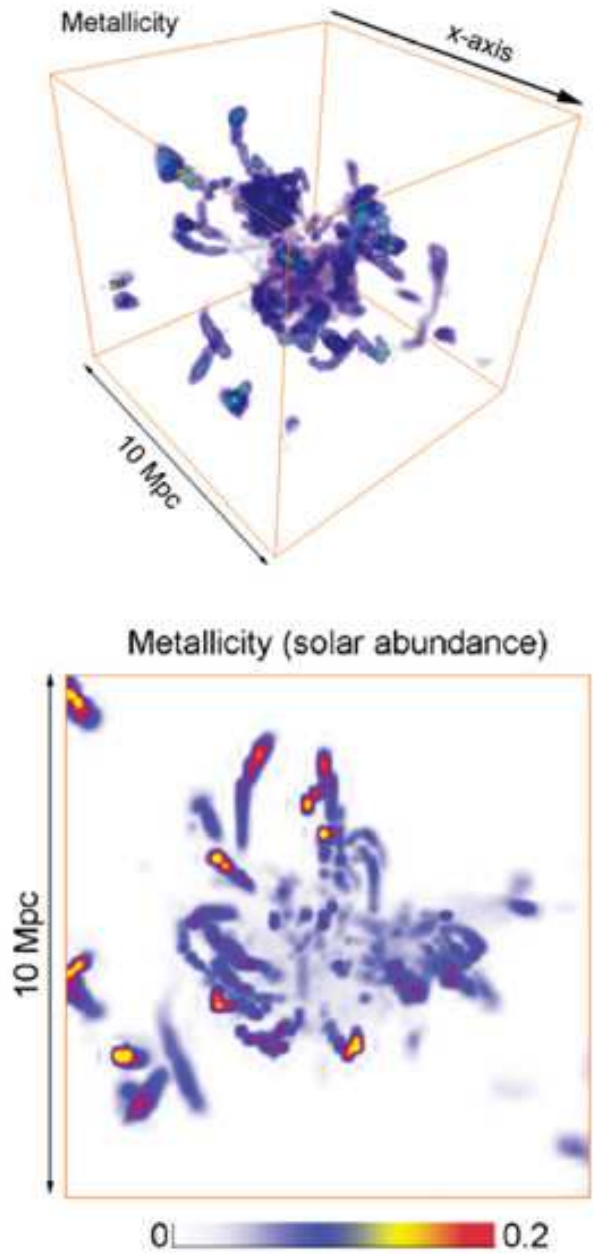,height=7in}
\caption {Metallicity map of a simulated galaxy
cluster. Top: 3D distribution. Bottom: metallicity map projected
along the x-axis.}
\end{figure}

\section*{Acknowledgments}
The authors acknowledge  the support of the Austrian Science
Foundation (FWF) through grant number P15868 and the Tyrolean
government.

\section*{References}
Breitschwerdt, D., McKenzie, J.F., V\"olk, H.J., 1991, A\&A, 245,
79\\
Colella, P., Woodward, P.R., 1984, J.Comput. Phys., 54,174\\
De Young, D.S., 1978, ApJ, 176,1\\
Fukazawa, Y., Makishima, K., Tamura, T. et al., 1972, ApJ, 176,
1\\
Gunn, J.E., Gott, J.R.III, 1972, ApJ, 176, 1\\
van Kampen, E., Jimenez, R., Peacock, J.A., 1999, MNRAS, 310, 43\\
Ruffert, M., 1992, A\&A, 265, 82\\
\end{document}